\documentclass[12pt]{article}

\usepackage{graphicx}

\begin{document}

\title{Phase transitions in some epidemic models defined on small-world
networks}
\date{}
\author{H. N. Agiza$^1$, A. S. Elgazzar$^2$ and S. A. Youssef$^1$ \\
$^1$ Mathematics Department, Faculty of Science\\
35516 Mansoura, Egypt\\
$^2$ Mathematics Department, Faculty of Education\\
45111 El-Arish, Egypt\\
E-mail: elgazzar@mans.edu.eg}
\maketitle

\begin{abstract}
Some modified versions of susceptible-infected-recovered-susceptible (SIRS)
model are defined on small-world networks. Latency, incubation and variable
susceptibility are included, separately. Phase transitions in these models
are studied. Then inhomogeneous models are introduced. In some cases, the
application of the models to small-world networks is shown to increase the
epidemic region.\newline
\end{abstract}

\section{ Introduction}

There are many mathematical models for epidemics [1-5]. Generally,
the population is classified into susceptible (S), infected (I)
and recovered (R) according to the state of each individual. The
SIRS model is proposed to describe the outbreaks of foot-and-mouth
disease (FMD) [1]. The model is generalized to include latency,
incubation and variable susceptibility [4]. They have studied
phase transitions in these models. Also it is shown that [5] a
ring vaccination programme is capable of eradicating FMD in SIRS
model defined on small-world networks (SWN).

The concept of SWN [2,3] is proposed to describe some real social
networks. Therefore SWN is used successfully to model several real
systems [5-8].

Here our aim is to study phase transitions in some modified
versions of SIRS model (including inhomogeneous mixing, latency,
incubation and variable susceptibility) defined on SWN.

The paper is organized as follows: In section 2, the concept of
SWN is explained. Phase transitions in some SIRS versions are
studied in section 3. In section 4, some generalized versions are
discussed. Some conclusions are summarized in Section 5.

\section{Small-world networks}

If one considers all human in the world are occupying the vertices
of a network, this social network has to satisfy two main
properties: clustering and small-world effect [9]. Clustering
means every one has a group of collaborators, some of them will
often be a collaborator by another individual. Small-world effect
means the average shortest vertex-to-vertex distance, $l$, is very
short compared with the size of the network $N$ (the total number
of vertices).

Regular lattices display the clustering property, because its clustering
coefficient is high. The clustering coefficient $(C)$ is defined as the
average fraction of pairs of neighbors of a vertex which are also neighbors
of each other. But regular lattices do not display the small-world effect,
because the distance $l$ increases as $N^{1/d}$ in $d$ dimensions.

For a random graph [10] with coordination number $z$, the total number of
vertices $N$ is given by

\[
N=z^{l},
\]
which gives,

\begin{equation}
l=\frac{\log \,N}{\log \,z}.
\end{equation}
The logarithmic increase with $N$ allows the distance $l$ to be very short
even for large $N$. Then random graphs display the small-world effect. But a
random graph does not satisfy the clustering property, because its
clustering coefficient is given by $C=z/N,\,$ this quantity goes to zero for
large $N$. Therefore both regular and random lattices are not good
descriptions for social networks.

A SWN consists of a regular one-dimensional (1-d) chain with periodic
boundary conditions. Each vertex is connected to its nearest-neighbors by
bonds. Some shortcutting bonds joining between some randomly chosen vertices
with probability $\phi $ are added. The probability $\phi $ is supposed to
be small in order to preserve the clustering property of the regular
lattice. The small-world effect is concluded as follows: for very small
lattice size $N$, it is less probable to find a shortcut, so the system
behaves like a regular lattice, and

\begin{equation}
l\propto N.
\end{equation}
When $N$ becomes large enough, more shortcuts are expected and the system
behaves as a random lattice i.e.

\begin{equation}
l\propto \log \,N.
\end{equation}
Consider this transition occurs at certain system size $\zeta $, then $l$
obeys a finite size scaling law as

\[
l=Nf\left(\frac{N}{\zeta }\right),
\]
where $f(x)$ is a universal scaling function, such that

\begin{equation}
f(x)=\left\{
\begin{array}{c}
{\rm{const.}}\;\;\;{\rm{if}}\; x\ll 1, \\
\frac{\log x}{x}\;\;\;\;\;\;{\rm{if}}\; x\gg 1.
\end{array}
\right.
\end{equation}
After some calculation using the renormalization group theory [9], one gets

\[
l=Nf(\phi N),
\]
for 1-d and considering the first-nearest neighbors only ($d=1$,
$k=1$), and

\begin{equation}
l=\frac{N}{k}f((\phi k)^{\frac{1}{d}}N),
\end{equation}
for general $k$-distance-nearest neighbors and any $d$. These
forms are valid only for $N\gg 1$ and $\phi \ll 1$. Then for $N\gg
1/\phi$,

\[
l\propto \log N.
\]
Then SWN are shown to combine both properties of social networks.
Also this structure combines between both local and non local
interactions which is observed in many real systems. Therefore the
concept of SWN is used in modelling several real systems [5-8].
Here our interest is restricted to apply the concept of SWN to
some epidemic models.

\section{Phase transitions in some epidemic models}

The population is classified into three classes: susceptible, infected and
recovered. Consider a function $s(i,t)$ represents the state of an
individual $i$ at time $t$, such that,

\[
s(i,t)=\left\{
\begin{array}{l}
-1\;\;\;{\rm{for\; I-individuals,}} \\
0\;\;\;\;\;\;{\rm{for\; S-individuals,}} \\
1\;\;\;\;\;\;{\rm{for\; R-individuals.}}
\end{array}
\right.
\]
The transitions between the states S, I and R occur according to the
following automata rules:

\begin{enumerate}
\item[ ]  Infection:

If $s(i,t)=0$ and ($s(i-1,t);s(i+1,t)$ or both $=-1$), then $s(i,t+1)=-1$
with probability $p_{1}$.

\item[ ]  Recovery:
\begin{equation}
{\rm{If}}\;s(i,t)=-1,\;{\rm{then}}\;s(i,t+1)=1.
\end{equation}

\item[ ]  Losing immunity:
\begin{equation}
{\rm{If}}\;s(i,t)=1,\;{\rm{then}}\;s(i,t+1)=0\;{\rm{with\;probability}}\;p_{
2}.
\end{equation}
\end{enumerate}
This model is approximated by the following set of differential
equations:

\begin{equation}
\frac{{\rm{d}}S}{{\rm{d}}t}=p_{2}R-p_{1}SI,
\end{equation}

\begin{equation}
\frac{{\rm{d}}I}{{\rm{d}}t}=p_{1}SI-I,
\end{equation}

\begin{equation}
\frac{{\rm{d}}R}{{\rm{d}}t}=I-p_{2}R
\end{equation}
This set of differential equations is a mean field approximation that
ignores the spatial structure of the lattice. It assumes a global
interacting system. But in reality a disease spreads locally with some non
local interactions. Also including some epidemic aspects like lower
susceptibility, incubation to the differential equations is very difficult.
On the other hand, it is allowed in lattice models [4,5].

To study the phase transition in this model, the probabilities
$p_{1}$ and $p_{2}$ are varied from $0$ to $1$ by step $0.01$. The
phase diagram is drawn as a relation between $p_{1}$ and $p_{2}$.
There are two limiting points: The first is when $p_{2}=0$
corresponding to the case of perfect immunization, and it is close
to ordinary percolation [11]. The second case is for $p_{2}=1$
representing the case of zero immunization, and this case is well
described by directed percolation [12]. For intermediate values of
$p_{2}$, there are no clear relation to the percolation theory.
The phase diagram is similar to damage spreading transitions [13],
where two phases appear: epidemic and non epidemic.

Ahmed and Agiza [4] have studied phase transitions in some
modified versions of 1-d SIRS model including inhomogeneous
mixing, latency, incubation and variable susceptibility. Here we
will generalize their work to SWN. The SWN used here is a 1-d
chain of size $1000$ with periodic boundary conditions. Shortcuts
are fixed beforehand with probability $\phi =0.05$ per bond. The
models evolve for $10000$ time steps. We will study four different
versions of SIRS model separately.

The first is the original SIRS itself. The automata rules are generalized to
include the shortcutting neighbors as follows:

\begin{enumerate}
\item[ ]  Infection: If $s(i,t)=0$ and $(s(i-1,t);\;s(i+1,t)$ or
$s(s_{c}(i),t)$
(if exists), at least $=-1$), then $s(i,t+1)=-1$ with probability
$p_{1},$ where $s_{c}(i)$ is the shortcutting neighbor of the
 $i$-th individual (if exists).
\end{enumerate}
Rules for both recovery and losing immunity are the same as the
original model. The phase diagram is shown in Fig. 1. It appears
that an epidemic phase occurs at $p_{1c}=0.61$. This value is
slightly less than that observed from the 1-d original model.

In some cases, an infection does not spread directly, but it needs some time
and suitable conditions to be transmitted. In order to model this
phenomenon, the population is classified into four states: susceptible,
infected, recovered and lower susceptible. Then the state function is
modified to:

\begin{equation}
s(i,t)=\left\{
\begin{array}{l}
-1\;\;\;{\rm{for\; I-individuals,}}\\
0\;\;\;\;\;\;{\rm{for\; S-individuals,}}\\
0.5\;\;\;{\rm{for \;lower \;susceptible \;individuals,}}\\
1\;\;\;\;\;\;{\rm{for\; R-individuals.}}
\end{array}
\right.
\end{equation}
A lower susceptible individual has immunity greater than a
susceptible individual; but smaller than a recovered one. The
model is defined on SWN. Consider $30\%$ of the population have a
lower susceptibility. Both infection and recovery rules are the
same as in the first case. The other rules are modified to:

\begin{enumerate}
\item[ ]  Losing immunity: if $s(i,t)=1$, then with probability $p_{2},$
\begin{equation}
s(i,t+1)=\left\{
\begin{array}{l}
0\;\;\;\;\;{\rm{with\;probability}}\;0.7, \\
0.5\;\;\;{\rm{with\;probability}}\;0.3.
\end{array}
\right.
\end{equation}

\item[ ]  Susceptibility:

If $s(i,t)=0.5$ and $(s(i-1,t);s(i+1,t)$ or $s(s_{c}(i),t)$ (if
exists), at least $=-1$), then $s(i,t+1)=0$.
\end{enumerate}
The phase diagram is given in Fig. 2. The epidemic phase occurs
at $p_{1c}=0.71$. The epidemic region is smaller than that of the
first case, because of the assumption that $30\%$ of the
population are not infected directly.

The third case, in some infectious diseases, a diseased individual can be
infecting but symptoms don't appear (incubation state). On the other hand,
an infected individual may not be infecting but still has the symptoms (i.e.
latent state). This case is called incubation-latent model. We define this
model on SWN as follows: The state function is modified to:

\begin{equation}
s(i,t)=\left\{
\begin{array}{l}
-2\;\;\;{\rm{represents \;incubation,}}\\
0\;\;\;\;\;\;{\rm{represents \;susceptibility,}}\\
1\;\;\;\;\;\;{\rm{represents \;recovery,}}\\
2\;\;\;\;\;\;{\rm{represents \;latency.}}
\end{array}
\right.
\end{equation}
The rule of losing immunity is the same as in the original model; but the
other rules are modified as follows:

\begin{enumerate}
\item[ ]  Incubation:

If $s(i,t)=0$ and ($s(i-1,t);s(i+1,t)$ or $s(s_{c}(i),t)$ (if
exists), at least $=-2$), then $s(i,t+1)=-2$ with probability
$p_{1}$.

\item[ ]  Recovery:

If $s(i,t)=2$, then $s(i,t+1)=1$.

\item[ ]  Latency:

If $s(i,t)=-2$, then $s(i,t+1)=2$.
\end{enumerate}
The results are shown in Fig. 3, the epidemic region extended
again with $p_{1c}=0.61$. This case is similar but not identical
to first case.

Fourth, an incubation both sick and infecting model is introduced.
In some diseases like Aids, a diseased person is sick and
infecting, so this model is called an incubation both sick and
infecting model. The state function $s(i,t)$ is defined as
follows:

\begin{equation}
s(i,t)=\left\{
\begin{array}{l}
-2\;\;\;{\rm{represents \;incubation,}}\\
-1\;\;\;{\rm{represents \;sick \;and \;infecting,}} \\
0\;\;\;\;\;\;{\rm{represents \;susceptibility,}}\\
1\;\;\;\;\;\;{\rm{represents \;recovery.}}
\end{array}
\right.
\end{equation}
The model is defined on SWN, and the automata rules become:

\begin{enumerate}
\item[ ]  Incubation:

If $s(i,t)=0$ and $(s(i-1,t)<0$; $s(i+1,t)<0$ or $s(s_{c}(i),t)<0$
(if exists), at least), then $s(i,t+1)=-2$ with probability
$p_{1}$.

\item[ ]  Sick-Infected:

If $s(i,t)=-2$, then $s(i,t+1)=-1$.
\end{enumerate}
Rules for both recovery and losing immunity are the same as in the
original model. The phase diagram is shown in Fig. 4, the
epidemic phase occurs at $p_{1c}=0.41$. The epidemic region is
extended more than the previous cases, because infection is
expected from both sick-infected and incubation individuals.

\section{Some generalizations}

Sometimes an infection is transmitted to some distant neighbors in
addition to the nearest neighbors. This interaction with the
distant neighbors is modelled by generalizing the automata rules,
discussed in the previous section, to include distant neighbors at
a distance $k$. This means $k=1$ gives the first-nearest
neighbors, $k=2$ gives the second-nearest neighbors in addition to
the first-nearest neighbors, and so on. The case $k=2$ is studied
for the four cases, and the results are summarized in table 1.
The epidemic phase increased significantly in the four cases. This
is expected, because the infection spreads faster than in the case
of $k=1$.

Generally, every individual has his/her own immune system that
differs significantly from the others. Thus the susceptibility
also differs from one to another. Also, in some cases the
probability of infection depends on the number of infected
neighbors. To model this behavior, a modified probability of
infection is considered. If $p_{1}$ is the probability of
infection due to one infected nearest neighbor, then $(1-$
$p_{1})^{m}$ is the probability of noninfection due to $m$
infected nearest neighbors. Then the modified probability of
infection [14] is

\begin{equation}
p_{1}^{*}=1-(1-p_{1})^{m},
\end{equation}
per unit of time. Beside the advantages of this form, it also
implies that the probability of infection for each individual is
not constant with time. Using $p_{1}^{*}$ instead of $p_{1}$
is introducing inhomogeneity that is one of the main aspects in
reality.

Inhomogeneous models are constructed for the four cases studied in
the previous section. The same conditions are applied. The results
are close to that of the models in SWN, but there are slight
differences from the results of the 1-d models. A comparison
between the results of regular lattice, SWN with $k=1$, SWN with
$k=2$ and the inhomogeneous models is given in Table 1.
\begin{center}
\begin{tabular}{|c|c|c|c|c|}
\hline $p_{1c}$ &
\begin{tabular}{c}
Regular lattice \\
$k=1$ [4]
\end{tabular}
& SWN $k=1$ & SWN $k=2$ &
\begin{tabular}{c}
Inhomogeneous \\
Models
\end{tabular}
\\ \hline
Case 1 & 0.68 & 0.61 & 0.38 & 0.60 \\ \hline Case 2 & 0.80 & 0.71
& 0.43 & 0.74 \\ \hline Case 3 & 0.67 & 0.61 & 0.40
& 0.59 \\ \hline Case 4 & 0.47 & 0.41 & 0.24 & 0.40 \\
\hline
\end{tabular}

Table 1: The critical value $p_{1c}$ at which the phase
transition occurs for all the studied models.
\end{center}

\section{Conclusions}

Phase transitions in some modified versions of SIRS model for
epidemics are studied using SWN with both $k=1$ and $k=2$. Also,
inhomogeneous models are introduced. Only the case of $k=2$ is
found to significantly affect the phase transitions in all models.
Just slight changes are found in the other cases.\newline
\newline
\textbf{Acknowledgements}\newline
\newline
We thank E. Ahmed for helpful discussions.

\newpage

\begin{figure}
\begin{center}
\includegraphics[angle=-90, width=0.5\textwidth]{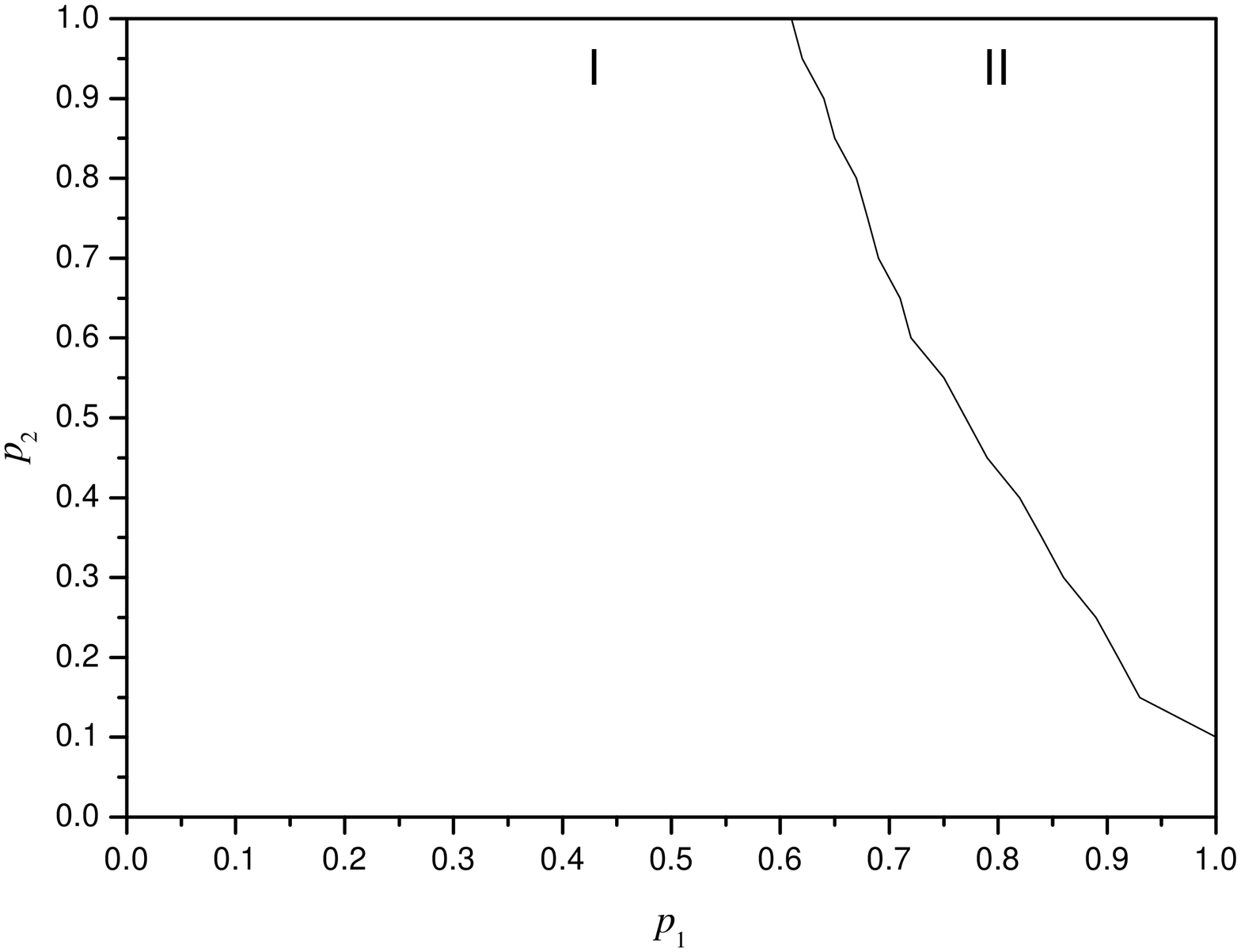}
\caption{Phase diagram for the SIRS model defined on SWN with
$N=1000$, $\phi =0.05$ and $k=1$. Tow phases appear nonepidemic
(I) and epidemic (II), and $p_{1c}=0.61$.}
\end{center}
\end{figure}

\begin{figure}
\begin{center}
\includegraphics[angle=-90, width=0.5\textwidth]{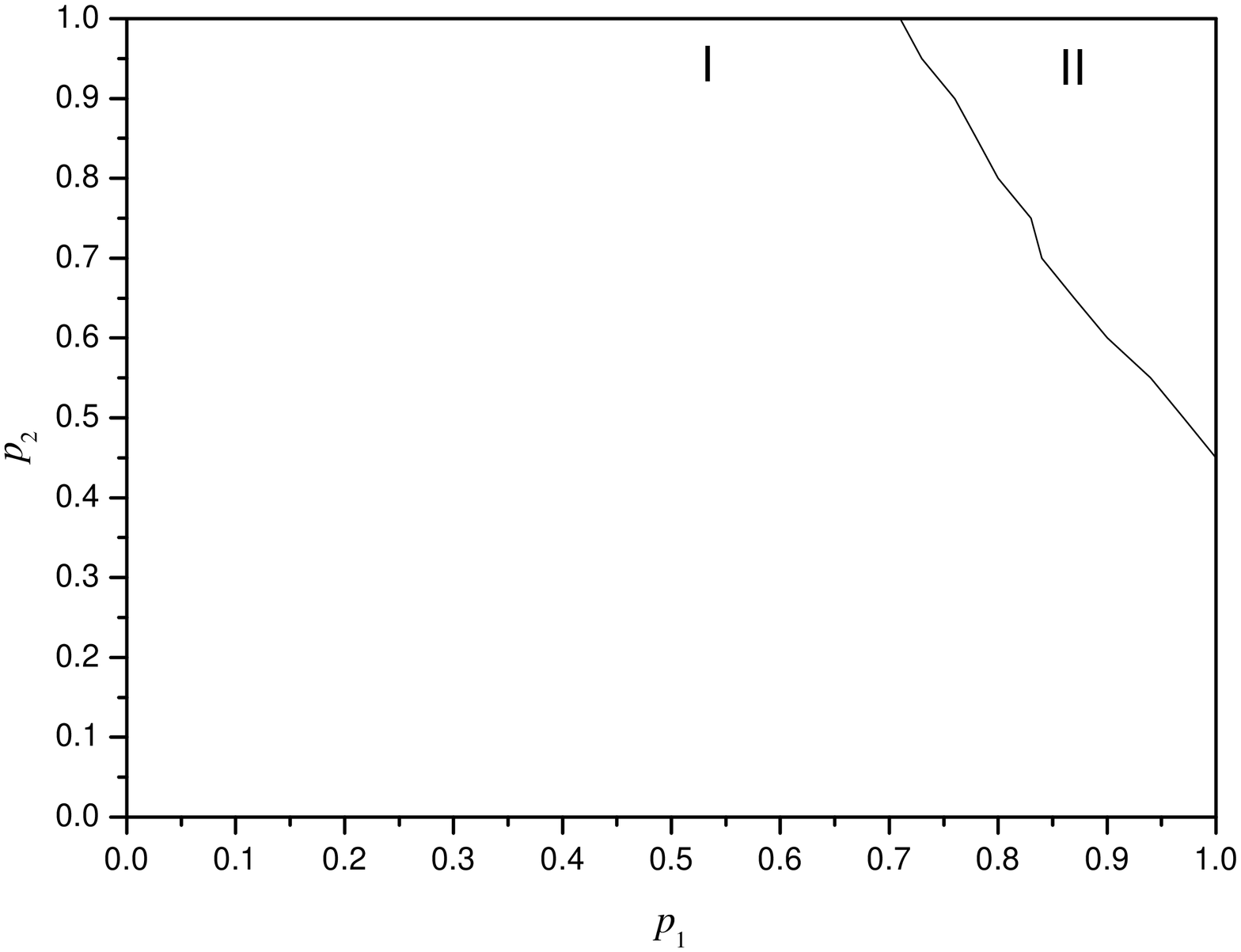}
\caption{Phase diagram for the SIRS model defined on SWN with
$N=1000$, $\phi =0.05$ and $k=1$. Lower susceptibility is
introduced. The critical value is $p_{1c}=0.71$.}
\end{center}
\end{figure}

\begin{figure}
\begin{center}
\includegraphics[angle=-90, width=0.5\textwidth]{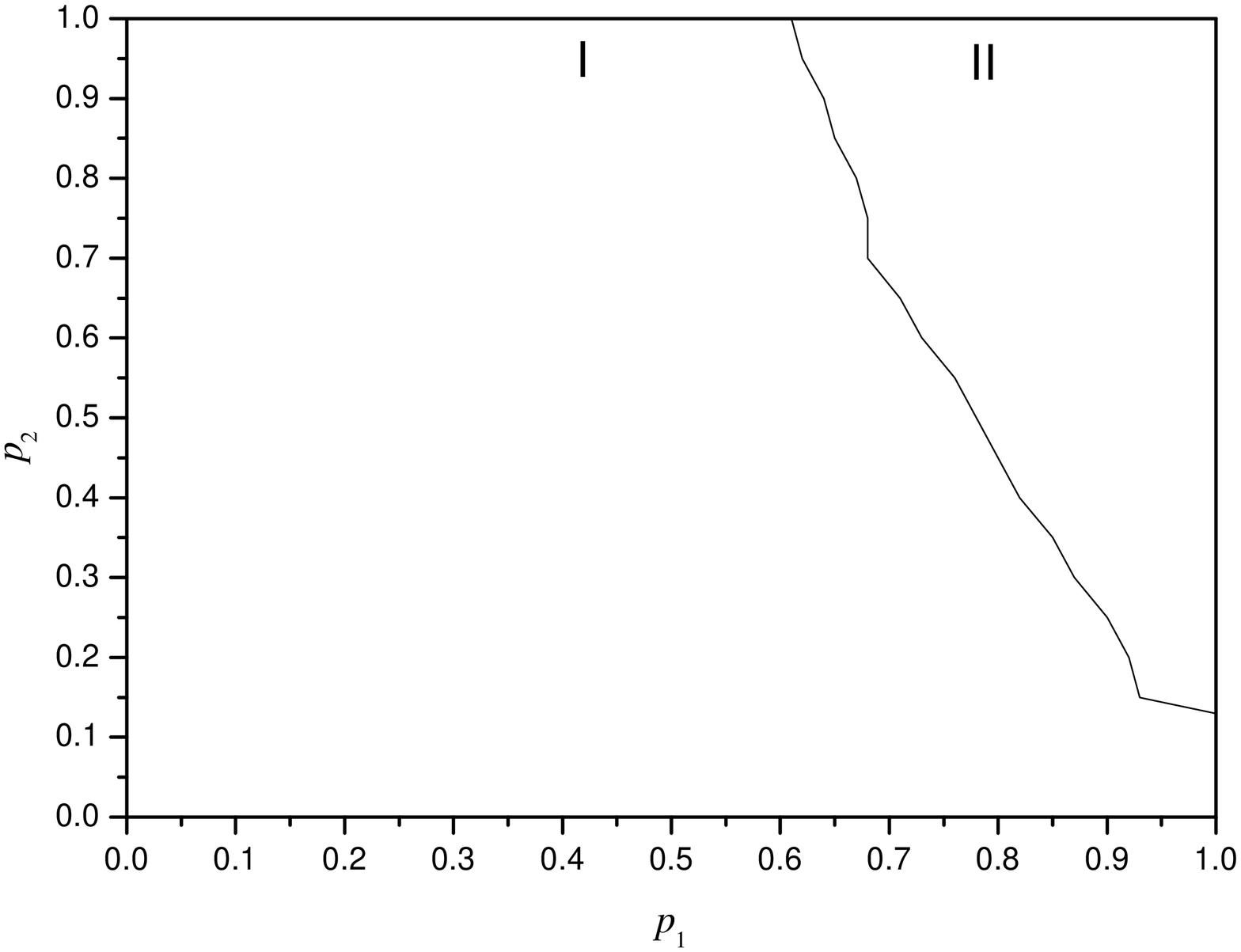}
\caption{Phase diagram for the third case, where an
incubation-latent model is defined on SWN with $N=1000$, $\phi
=0.05$ and $k=1$. The phase transition occurs at $p_{1c}=0.61$.}
\end{center}
\end{figure}

\begin{figure}
\begin{center}
\includegraphics[angle=-90, width=0.5\textwidth]{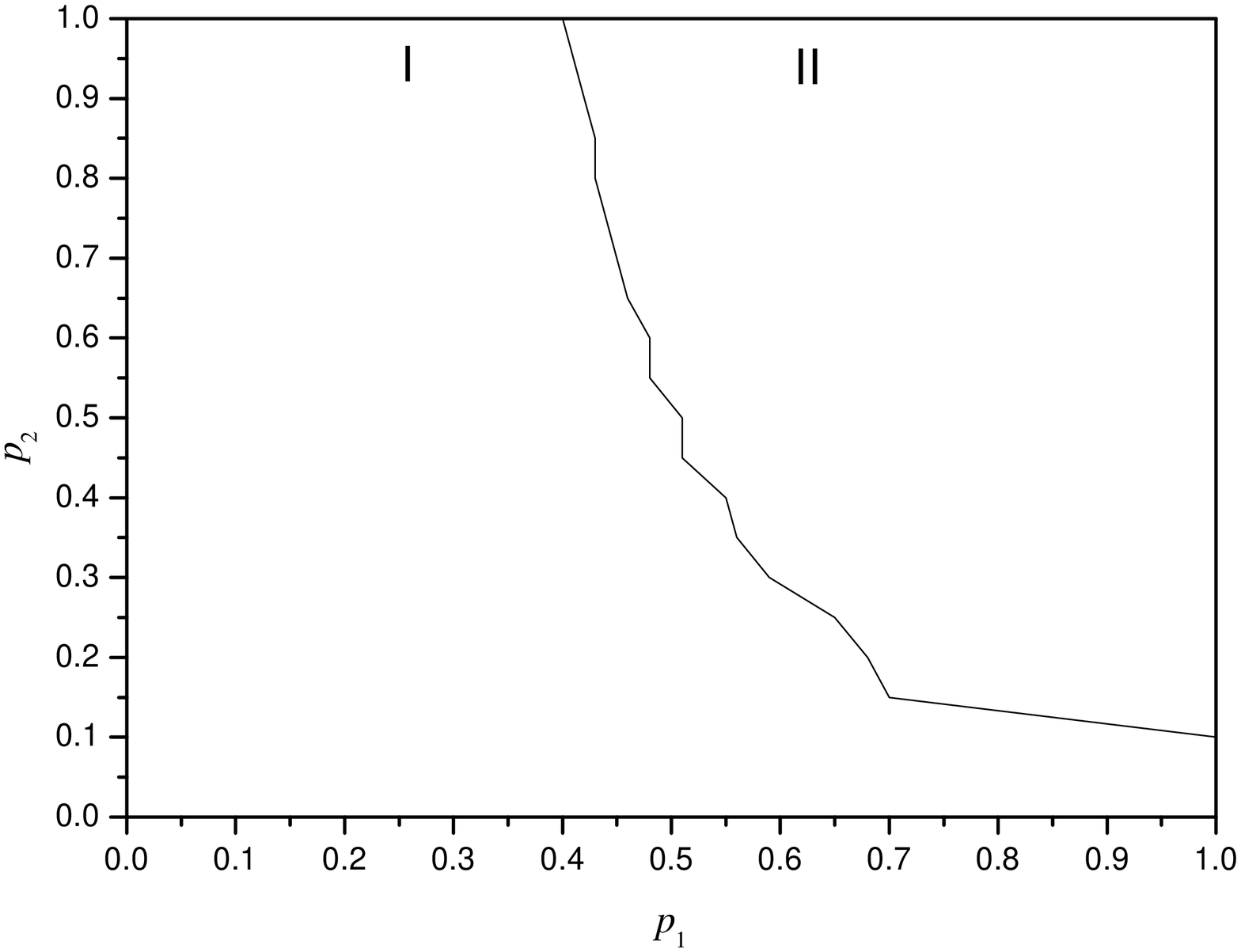}
\caption{Phase diagram for a modified version of SIRS model
including the concept of incubation both sick and infecting. The
model is defined on SWN with $N=1000$, $\phi =0.05$ and $k=1$. A
phase transition from nonepidemic to epidemic phases is observed
at $p_{1c}=0.41$.}
\end{center}
\end{figure}

\end{document}